\pdfoutput=1
\documentclass[sigconf,screen]{acmart}

\usepackage{bbm}
\usepackage{mathtools}
\usepackage[ruled, algo2e, vlined]{algorithm2e}

\usepackage{url}

\usepackage{tcolorbox}

\allowdisplaybreaks

\AtBeginDocument{%
  \providecommand\BibTeX{{%
    \normalfont B\kern-0.5em{\scshape i\kern-0.25em b}\kern-0.8em\TeX}}}

\copyrightyear{2022}
\acmYear{2022}
\setcopyright{acmcopyright}\acmConference[CIKM '22]{Proceedings of the 31st ACM International Conference on Information and Knowledge Management}{October 17--21, 2022}{Atlanta, GA, USA}
\acmBooktitle{Proceedings of the 31st ACM International Conference on Information and Knowledge Management (CIKM '22), October 17--21, 2022, Atlanta, GA, USA}
\acmPrice{15.00}
\acmDOI{10.1145/3511808.3557716}
\acmISBN{978-1-4503-9236-5/22/10}

\begin{document}

\title{Twin Papers: A Simple Framework of Causal Inference for Citations via Coupling}


\author{Ryoma Sato}
\email{r.sato@ml.ist.i.kyoto-u.ac.jp}
\affiliation{%
  \institution{Kyoto University / RIKEN AIP}
  \city{Kyoto}
  \country{Japan}
}

\author{Makoto Yamada}
\email{myamada@i.kyoto-u.ac.jp}
\affiliation{%
  \institution{Kyoto University / RIKEN AIP}
  \city{Kyoto}
  \country{Japan}
}

\author{Hisashi Kashima}
\email{kashima@i.kyoto-u.ac.jp}
\affiliation{%
  \institution{Kyoto University / RIKEN AIP}
  \city{Kyoto}
  \country{Japan}
}

\begin{abstract}
The research process includes many decisions, e.g., how to entitle and where to publish the paper. In this paper, we introduce a general framework for investigating the effects of such decisions. The main difficulty in investigating the effects is that we need to know counterfactual results, which are not available in reality. The key insight of our framework is inspired by the existing counterfactual analysis using twins, where the researchers regard twins as counterfactual units. The proposed framework regards a pair of papers that cite each other as twins. Such papers tend to be parallel works, on similar topics, and in similar communities. We investigate twin papers that adopted different decisions, observe the progress of the research impact brought by these studies, and estimate the effect of decisions by the difference in the impacts of these studies. We release our code and data, which we believe are highly beneficial owing to the scarcity of the dataset on counterfactual studies.
\end{abstract}



\maketitle

\section{Introduction}

It has been studied for a long time what aspects of research processes affect the number of citations \cite{davis2008open, onodera2015factors, tahamtan2018core, sato2022poincare}. Namely the publication venues \cite{yan2011citation, traag2021inferring, sato2022poincare, xiao2016modeling, xiao2020discovering}, authors \cite{haslam2008makes, yan2011citation, xiao2016modeling}, titles \cite{jamali2011article, annalingam2014determinants, stremersch2015unraveling, buter2011non, subotic2014short}, references \cite{vieira2010citations, webster2009hot}, and topological features \cite{yu2012citation, davletov2014high} have been considered as the cause of citations. For example, \citet{yan2011citation} argue that the authors' expertise and venue impact are important factors, and \citet{paiva2012articles} found that articles with short titles describing the results were cited more often.

Except for a notable exception of \citet{davis2008open}, who conducted a randomized control trial for investigating the impact of the choice of open access, most studies are based on observational studies. This is primarily because intervening research processes, e.g., by randomly changing publication venues or titles, may cause adverse impacts on the researchers' careers. For this reason, most of the existing studies investigate only correlations. Although some studies \cite{falagas2013impact, traag2021inferring, sato2022poincare} tried to find causal relations, they assumed specific statistical models and covariates. However, in general, the choice of covariates is not straightforward and crucially affects the results of analysis \cite{vanderweele2019principles}. In this paper, we propose a simple framework for adjusting confounders and thereby enabling us to find causal relationships in research processes and citations. Our framework can also be used for screening important factors before manual analysis.

\begin{tcolorbox}[colframe=gray!20,colback=gray!20,sharp corners]
\textbf{Reproducibility}: Our code and the list of twin papers are available at \url{https://github.com/joisino/twinpaper}.
\end{tcolorbox}

\section{Our Approach}

Let us consider a binary decision in the research process (e.g., whether to use a colon in the title, or publishing the paper in CIKM or SIGIR). We use whether to use a colon in the title as a running example. The goal of this study is to investigate whether a colon in the title increases the number of citations, and if any, how many citations. We consider a potential outcome framework for causal estimation, where the outcome is defined as the base-$2$ logarithm (instead of the raw value because of its broad dynamic range) of the number of citations a paper receives after a certain period. We say a paper $x$ receives a treatment if a colon is used in the title of $x$. There are two possible outcomes $Y_x(1)$ and $Y_x(0)$, the outcome value if the paper receives (resp. does not receive) the treatment. The quantity we want to estimate is: \begin{align}
    \text{ITE}_x &\stackrel{\text{def}}{=} Y_x(1) - Y_x(0), \\
    \text{ATE} &\stackrel{\text{def}}{=} \mathbb{E}_{x \sim p(x)}[\text{ITE}_x] = \mathbb{E}_{x \sim p(x)}[Y_x(1) - Y_x(0)],
\end{align}
i.e., how much the treatment increases the outcome in expectation. However, the critical problem is that we can observe only one of the two outcomes because we cannot publish the same paper with and without a colon simultaneously. Let $Y_x^F$ and $Y_x^C$ be the factual and counterfactual outcome values, respectively, i.e., if paper $x$ is published with a colon in the title, $Y_x^F = Y_x(1), Y_x^C = Y_x(0)$, and otherwise, $Y_x^F = Y_x(0), Y_x^C = Y_x(1)$. One cannot obtain $\text{ITE}_x$ because $Y_x^C$ is not observable.

\begin{figure*}[t]
 \begin{minipage}{0.33\hsize}
  \centering
    \includegraphics[width=\hsize]{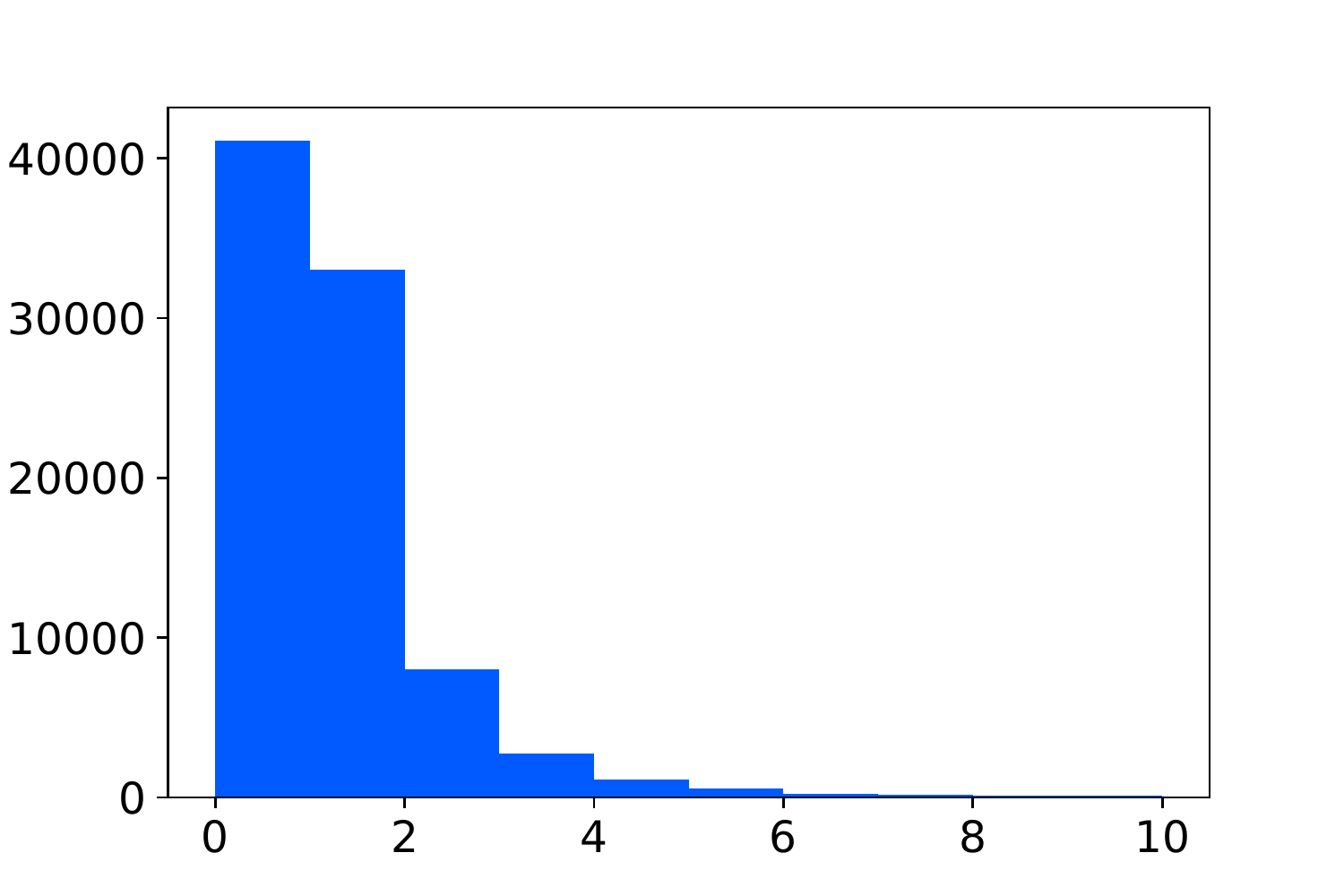}
    (a) Difference of publication years.
 \end{minipage}
 \begin{minipage}{0.33\hsize}
  \centering
    \includegraphics[width=\hsize]{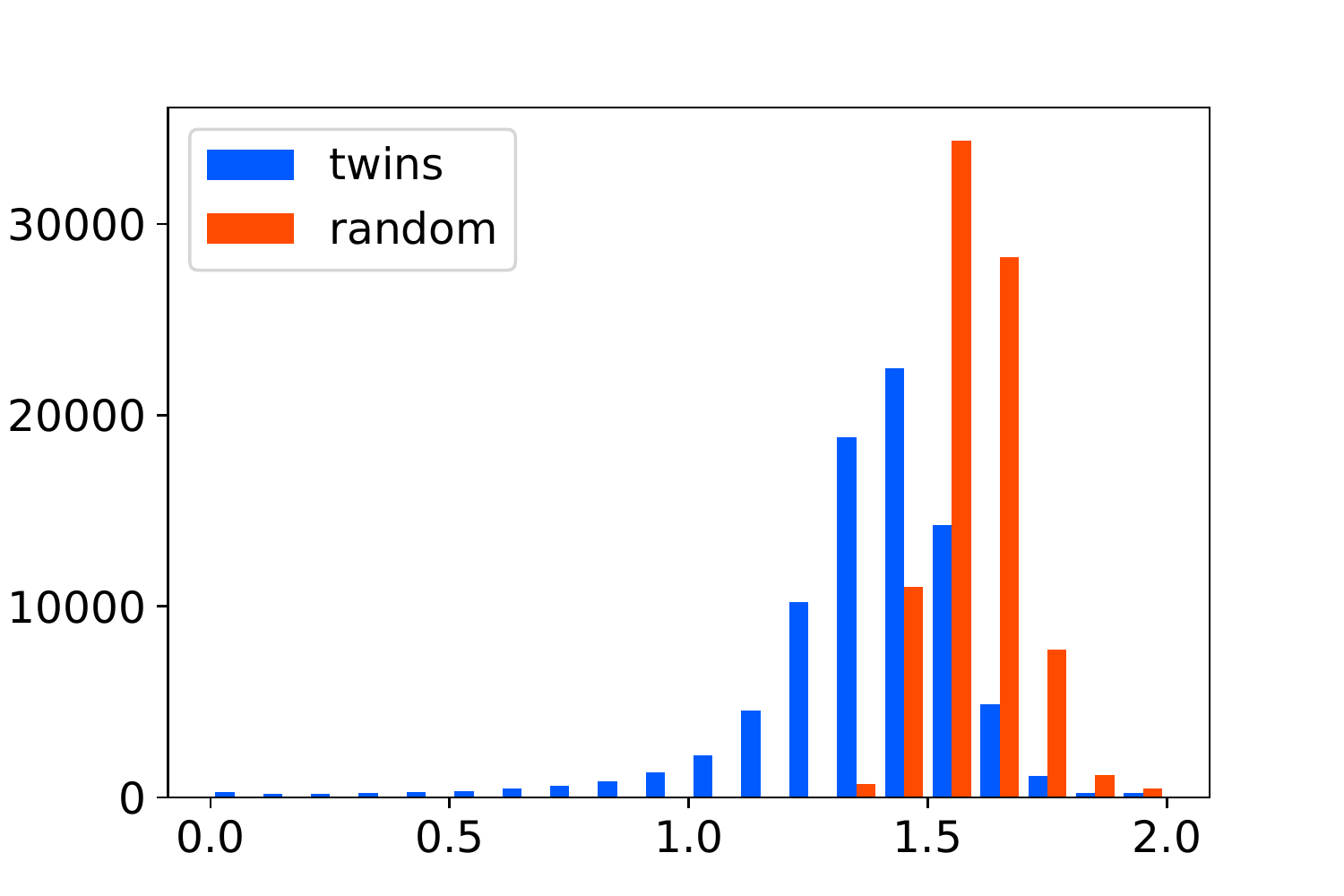}
    (b) Distances of abstracts.
 \end{minipage}
  \begin{minipage}{0.33\hsize}
  \centering
    \includegraphics[width=\hsize]{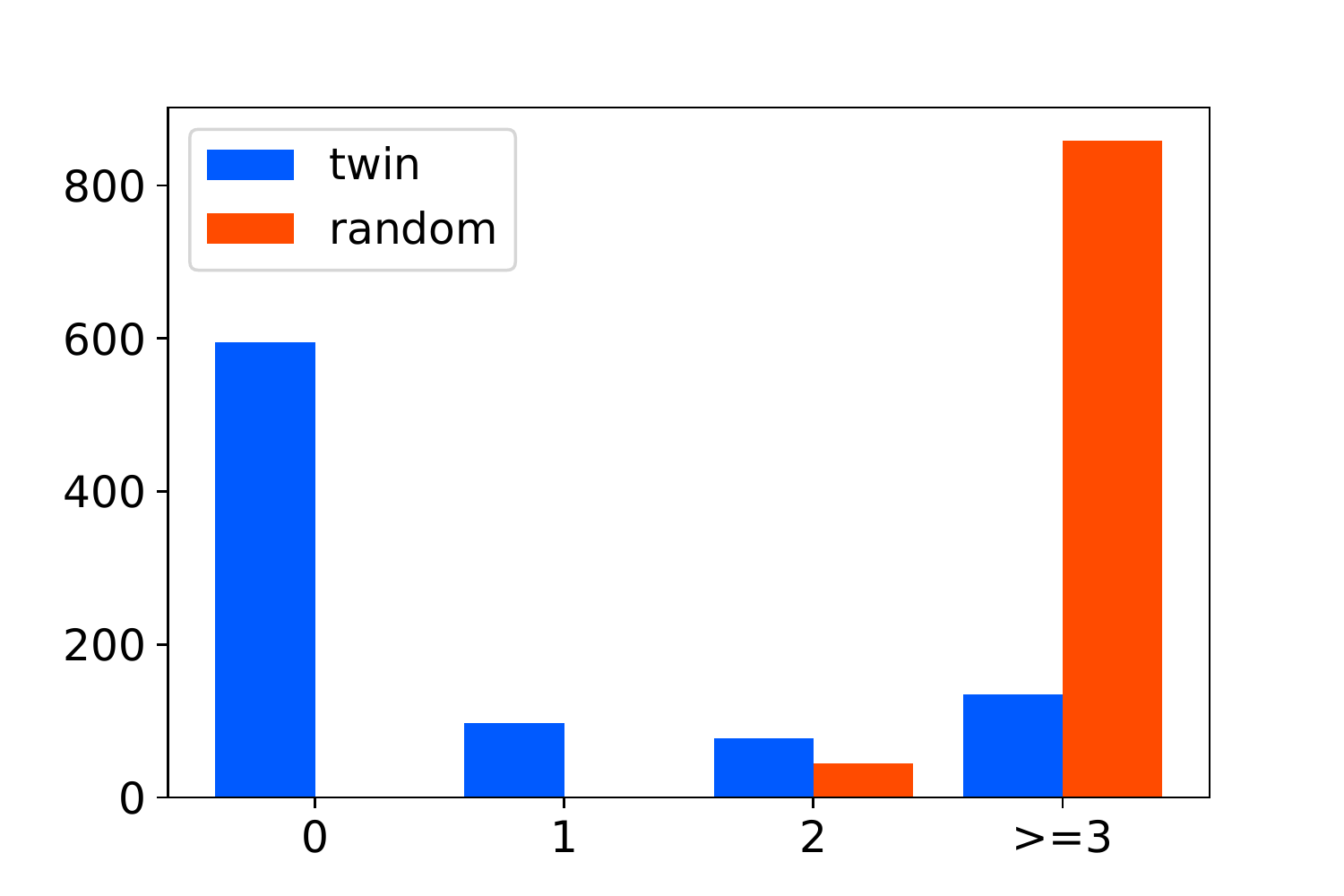}
    (c) Distances of twins on the collaboration network.
 \end{minipage}
 \vspace{-0.1in}
\caption{(a) Most twin papers are parallel works. (b) Most twins have similar topics. (c) Most twins belong to close communities. These results corroborate the assumptions of the twin paper framework.}
 \vspace{-0.1in}
 \label{fig: assumptions}
\end{figure*}

\section{Method}

Our proposed framework is inspired by the causal inference framework based on twins \cite{mcgue2010causal} in the medical and psychological domains.
The key insight of our proposed framework, twin papers, is that we can roughly regard a pair of papers that cite each other as counterfactual units. We call such a pair of papers twins. The rationale behind this definition is that twin papers tend to be (1) parallel works, (2) on similar topics, and (3) in close communities, which we will empirically show in the experiments. Therefore, twin papers can adjust many, if not all, confounders, including observable and unobservable ones. If the numbers of citations the twin papers receive are different, we can investigate what made the difference. Suppose a paper $x$ was published with a colon in the title and has a twin paper $y$ which was published without a colon. Then, we can estimate ITE by
\[ \widetilde{\text{ITE}} = Y_x^F - Y_y^F. \]
\sloppy This value can be computed solely from factual values. However, this estimate is noisy and has a high variance. Therefore, we consider the average effect, i.e., ATE. Let $\mathcal{D}^\text{colon/no colon} = \{(s, t) \mid s \text{ and } t \text{ are twins, and } s \text{ has a colon, } t \text{ has no colons} \}$. Then, ATE can be estimated by
\begin{align} \label{eq: ate}
\widetilde{\text{ATE}}^\text{colon/no colon} = \frac{1}{|\mathcal{D}^\text{colon/no colon}|} \sum_{(s, t) \in \mathcal{D}^\text{colon/no colon}} Y_s^F - Y_t^F.
\end{align}
This value can be computed solely from factual values.

We gather twin papers from the dblp dataset \cite{tang2008arnetminer}. There are $87,396$ twins in total, which are available in \url{https://github.com/joisino/twinpaper}.

\begin{table*}[tb]
    \centering
    \caption{Twin papers that are not parallel works. They are due to noise of the dataset.}
    \vspace{-0.15in}
    \scalebox{0.66}{
    \begin{tabular}{ll} \toprule
        Paper A & paper B \\ \midrule
A note on fast Fourier transforms for nonequispaced grids (1998) & Fast Fourier transforms for nonequispaced data (2018) \\
Analysis of an Off-Line Intrusion Detection System: A Case Study in Multi-Objective Genetic Algorithms (2005) & IMPROVED OFF-LINE INTRUSION DETECTION USING A GENETIC ALGORITHM (2016) \\
Formula-Dependent Equivalence for Compositional CTL Model Checking (1994) & Formula-Dependent Equivalence for Compositional CTL Model Checking (2002) \\ \bottomrule
    \end{tabular}
    }
    \label{tab: year_diff_anormaly}
    \vspace{-0.1in}
\end{table*}

\section{Illustrative Example}

To illustrate the benefit of twin papers, we create a subset of the dataset that contains only papers published in Symposium on the Theory of Computing (STOC), Symposium on Foundations of Computer Science (FOCS), Neural Information Processing Systems (NeurIPS), and International Conference on Machine Learning (ICML). STOC and FOCS are prestigious venues in theoretical computer science, and NeurIPS and ICML are prestigious venues in machine learning. As an example, we consider if adding a word ``learning'' in the title has a positive effect on the impact. We consider a paper with ``learning'' in the title to be treated and that without it is controlled. Intuitively, just changing the title of this paper to ``Twin Papers: A Simple Learning Framework...,'' would not change the number of citations much. Therefore, we expect the effect is small or zero. A naive approach to estimating the effect with observational data is,
\begin{align*}
&\widetilde{ATE}_\text{observational} \\
&=\frac{1}{|\{i \text{ is treated}\}|} \sum_{i \text{ is treated}} Y_i^F - \frac{1}{|\{i \text{ is controlled}\}|} \sum_{i \text{ is controlled}} Y_i^F.
\end{align*}
However, there is a selection bias because papers in NeurIPS and ICML tend to have ``learning'' in the title. In fact, $\widetilde{ATE}_\text{observational} = 0.132$, which indicates that the treatment has a positive effect. This result just reflects the fact that NeurIPS and ICML papers tend to receive more citations than STOC and FOCS papers. By contrast, if we use twin papers and the proposed estimator (i.e., \eqref{eq: ate}), $\widetilde{ATE} = -0.017$, which indicates the treatment has no effects.

\section{Confirming Assumptions} \label{sec: assumptions}

\subsection{Twins Tend to Be Parallel Works} \label{sec: parallel}

Figure \ref{fig: assumptions} (a) shows the histogram of the differences of publication years between twin papers. This indicates that $84.8$ percent of twin pairs are published in the same or the next year. However, some twin papers are published in different periods. We investigate the cause of this phenomenon. We draw random twin pairs whose publication years are different by more than five years and show them in Table \ref{tab: year_diff_anormaly}. The difference in the first example is as many as twenty years. We found out that this is because there is a paper with the same title as ``Fast Fourier transforms for nonequispaced data'' published in 1993. The dblp dataset confused these papers, maybe in the data processing process, and spurious twins are detected. Other examples were caused due to similar reasons. Overall, twins that are not parallel works are spurious twins caused by noise in the dataset. Optionally, we can remove such pairs by preprocessing, e.g., thresholding the difference of publication years by one or two. We use the original data in the following analysis because such cases are rare, as shown in Figure \ref{fig: assumptions}, and do not affect the results much.

\subsection{Twins Tend to Be on the Same Topic}

We compute L1-normalized L1 bag-of-words distances \cite{sato2022reeval} of the abstracts of twin papers. We then compute the same distances for random pairs. The histograms in Figure \ref{fig: assumptions} (b) show that twin papers tend to have similar abstracts. This indicates that twins tend to be on the same or similar topics.

\subsection{Twins Tend to Be in the Same Community}

We build a collaboration network, where a node is a researcher, and an edge indicates that two researchers have collaborated, using the dblp dataset. Researchers close in the collaboration networks are considered to be in the same research community. We compute the distance of two papers A and B as the minimum distance between the authors of paper A and the authors of paper B in the collaboration network. Figure \ref{fig: assumptions} (c) shows that the authors of twin papers tend to be close in the collaboration network.

\begin{table}[tb]
    \centering
    \caption{ATEs for the contents and papers. Intuitively, this table reads that if we make the treatment in the first column, the paper has more impact by the amount in the third column. The second column shows the number of twins from the dataset, e.g., $|\mathcal{D}^\text{colon/no colon}|$ in the colon case.}
    \vspace{-0.15in}
    \begin{tabular}{lrr} \toprule
        Treatment & $|\mathcal{D}|$ & $\widetilde{\text{ATE}}$ \\ \midrule
Including a Colon in the Title & 21080 & 0.356 \\
Lengthening the Title & 84970 & -0.126 \\
Lengthening the reference & 81857 & 0.710 \\ 
Lengthening the abstract & 82917 & 0.248 \\
Lengthening the paper & 65730 & 0.630 \\ 
Self citation & 10582 & 1.30 \\ \bottomrule
    \end{tabular}
    \label{tab: content}
    \vspace{-0.1in}
\end{table}

\begin{table*}[tb]
    \centering
    \caption{ATEs for the publication venues. We choose treatments so that ATE is positive. Therefore, intuitively, this table reads that the venue in the first column is better than that in the second column by the amount in the fourth column.}
    \vspace{-0.15in}
    \begin{tabular}{llrr} \toprule
        Treatment (a) & Control (b) & $|\mathcal{D}^{a/b}|$ & $\widetilde{\text{ATE}}$ \\ \midrule
Journal of Cognitive Neuroscience & NeuroImage & 817 & 0.539 \\
IEEE Transactions on Information Theory & International Symposium on Information Theory & 459 & 1.93 \\
Neural Computation & IEEE Transactions on Neural Networks & 216 & 0.632 \\
Neural Computation & Neural Networks & 199 & 0.76 \\
Symposium on the Theory of Computing & Foundations of Computer Science & 182 & 0.252 \\
IEEE Transactions on Signal Processing & International Conference on Acoustics, Speech, and Signal Processing & 178 & 2.78 \\
Neural Computation & Neurocomputing & 153 & 2.41 \\
Journal of Economic Theory & Games and Economic Behavior & 125 & 0.668 \\
IEEE ACM Transactions on Networking & International Conference on Computer Communications & 110 & 1.29 \\
Symposium on the Theory of Computing & Symposium on Discrete Algorithms & 105 & 0.693 \\ \bottomrule
    \end{tabular}
    \label{tab: venue}
    \vspace{-0.1in}
\end{table*}

\begin{table}[tb]
    \centering
    \caption{ATEs for the combination of treatments. Intuitively, this table reads that if we make the treatments in the first and second columns, the paper has more impact by the amount in the third column.}
    \vspace{-0.15in}
    \begin{tabular}{llrrr} \toprule
        Treatment A & Treatment B & $|\mathcal{D}|$ & $\widetilde{\text{ATE}}$ \\ \midrule
Lengthen the reference & Lengthen the paper & 41546 & 1.04 \\ 
Lengthen the reference & Self citation & 6473 & 1.75 \\ 
Lengthen the paper & Self citation & 5018 & 1.66 \\ 
\bottomrule
    \end{tabular}
    \label{tab: combination}
    \vspace{-0.2in}
\end{table}

\section{Analysis with Twins}

\textbf{Contents and Styles.} As discussed in the informetrics literature \cite{jamali2011article, annalingam2014determinants, stremersch2015unraveling, buter2011non, subotic2014short}, the contents and style of the paper may affect the number of citations of the paper. We investigate six treatments quantitatively.

First, some researchers name their method and put it at the beginning of the paper with a colon. For example, this paper starts with ``Twin Papers: A Simple...'' Such paper titles are catchy and may provide more chances of clicks. The second row of Table \ref{tab: content} shows that including a colon in the title slightly improves the number of citations. This finding is consistent with the findings of \citet{buter2011non}, who reported that adding a colon in the title had a positive effect on the number of citations. We note that the impact of a colon has been controversial \cite{hartley2007planning, jamali2011article, paiva2012articles} and may depend on domains \cite{buter2011non}. Our analysis is done with computer science papers (i.e., dblp papers), and the conclusion may not be appropriate for other domains. However, we emphasize that our framework is general and can be applied to other domains if used with other datasets.

Second, the lengths of paper titles vary from paper to paper. Short titles are easy to understand and provide strong impressions, whereas long titles have more chances to be caught in the eye and to be listed in search engines. We investigate which is better quantitatively. For each pair of twins, we consider that the one with the shorter title is treated and the other is controlled. We remove the pairs with the same title lengths. The first row of Table \ref{tab: content} shows that shortening the title is slightly better, which is consistent with \citet{ayres2000determinants}, but the effect is small. The effect of longer titles has been a controversial topic in the informetrics domain \cite{stremersch2007quest, jacques2010impact, habibzadeh2010shorter, jamali2011article, subotic2014short}, and sometimes the opposite effects have been confirmed \cite{jacques2010impact, habibzadeh2010shorter}. The small effect observed in this analysis is consistent with the literature.

A paper with more references may have more chances of reverse lookups of references. The third row of Table \ref{tab: content} shows that lengthening the reference has a moderately positive effect on the number of citations. This result is consistent with the findings of \citet{haslam2008makes} and \citet{onodera2015factors}.

Then, we consider the length of the abstract. A paper with a longer abstract has more chance of being searched. The fourth row of Table \ref{tab: content} shows that longer abstract has a slightly positive effect.

Next, we investigate the length of the paper. A longer paper is considered to have more content and evidence. Besides, the longer the paper is, the more chances it has to be caught by search queries. On the other hand, readers may be reluctant to read too long papers. The fifth row of Table \ref{tab: content} shows that lengthening the paper has a moderately positive effect. This result is consistent with the findings of \citet{falagas2013impact}.

Finally, self citation is a common strategy to increase the number of citations \cite{aksnes2003macro, fowler2007does}. Self citations do not only increase the number of citations directly but also improve the exposure. Furthermore, many scholarly search engines such as Google Scholar and Semantic Scholar provide citation numbers in the search results, and the increase of citation numbers will increase the chances of clicks. We consider a paper is treated if the paper is cited by a paper that has at least one common author. The sixth row of Table \ref{tab: content} shows that a self citation has a strong positive effect on the number of citations. This is consistent with the findings of \citet{fowler2007does}.

\vspace{0.1in}
\noindent \textbf{Priority.} Although twin papers are parallel works, their publication dates are slightly different. We consider the one published earlier is treated in this analysis. Surprisingly, the estimated ATE was $-0.187$, which means that earlier publications receive slightly fewer citations. We hypothesize that this is because the quality of later publications is better. At least, this result indicates that hurrying to publish does not benefit in the long run.

\vspace{0.1in}
\noindent \textbf{Venue.} Publishing a paper in a conference or journal makes the paper known in the community and has an effect on the impact of the paper in the community. As each venue has different readers and participants, different venues may have different effects. We investigate the impact of the choice of venue in this section.
First, for each pair $(a, b)$ of venues, we construct $\mathcal{D}^{a/b} = \{(s, t) \mid s \text{ and } t \text{ are twin, } s \text{ is published in } a \text{, } t \text{ is published in } b \}$. We estimate ATE of publishing in $a$ over publishing in $b$ based on Eq. \eqref{eq: ate}. Table \ref{tab: venue} shows the results for the venues with the $10$ largest $|\mathcal{D}^{a/b}|$. We can observe that publishing in IEEE Transaction on Signal Processing provokes many citations compared to publishing in International Conference on Acoustics, Speech, and Signal Processing (ICASSP). In addition, publishing in Symposium on the Theory of Computing (STOC) is comparable to publishing in Foundations of Computer Science (FOCS), although STOC is slightly better. By contrast, STOC is clearly better than Symposium on Discrete Algorithms (SODA).

\subsection{Are the effects additive?}

Suppose treatment A doubles the number of citations and treatment B doubles the number of citations. Then, if we adopt both treatments A and B, will the number of citations quadruple? Note that as we measure the effect in a log domain, if the effect is additive, the number of citations is multiplicative. Table \ref{tab: combination} shows that the effect is sub-additive. For instance, the effects of lengthening the reference and paper are $0.710$ and $0.630$, respectively, according to Table \ref{tab: content}, and the effect of both treatments is $1.04 < 0.710 + 0.630$. However, combining several positive treatments does have positive effects and is better than a single treatment. It should be noted that the linear models adopted in previous research \cite{habibzadeh2010shorter, falagas2013impact, abramo2019predicting} cannot handle this kind of nonlinearity.

\section{Discussion}

\subsection{Other Confounders}
We confirmed that twin papers adjusted three conditions in Section \ref{sec: assumptions}. We argue that much more conditions are adjusted by twin papers. For example, the research problems they tackle are considered to be the same or similar. Besides, we hypothesize that the qualities of papers would also be adjusted to some extent, if not totally, because too low-quality papers are unlikely to be cited. Importantly, the quality of a paper is difficult to quantify, and thus we cannot numerically validate this hypothesis. We argue that the ability to control such unobservable/unquantifiable confounders is the strength of twin papers because other methods such as multivariate analysis cannot handle them.

\subsection{Limitations}
First, twin papers do not necessarily control \emph{all} confounding factors. For example, if authors decide the venue, and after that, they decide to add a colon in the title following the custom of the venue, then, the choice of the venue becomes a confounding factor. In this case, one needs to adjust the confounding factor using auxiliary features. We stress that our framework is general and can be combined with other adjustment methods such as multivariate analysis and stratified analysis, and importantly, the strength of twin papers is that it can adjust many, if not all, factors with a simple procedure.

Second, strictly speaking, twin papers are not \emph{true} counterfactual results. In reality, if two papers have similar topics and cite each other, the research impacts of these papers affect one another. This limitation is common with the original study on twins. However, the twin paper framework is much less sensitive to biases than previous studies using random samples from observational data. Combining our framework with manual analysis, e.g., multivariate analysis and stratified analysis, will further mitigate this problem.

The third limitation is that twins are rare in some domains. We found that the data mining domain had few twins. We hypothesize that this is because many data mining conferences prohibit submitting papers to arXiv during submission, and it hinders authors from finding concurrent papers on the same topic.

\section{Conclusion}

In this paper, we proposed a simple framework for investigating the effect of the decisions in research processes. We empirically confirm that twin papers are published under similar conditions, and conduct several case studies on the effects on the contents of the paper and publication venues.

\begin{acks}
This work was supported by JSPS KAKENHI GrantNumber 21J22490 and JST CREST Grant Number JPMJCR21D1.
\end{acks}


\bibliographystyle{plainnat}
\bibliography{sample-base}










\end{document}